\documentclass{article}

\usepackage{arxiv}

\usepackage[utf8]{inputenc} 
\usepackage[T1]{fontenc}    
\usepackage{hyperref}       
\usepackage{url}            
\usepackage{booktabs}       
\usepackage{amsfonts}       
\usepackage{nicefrac}       
\usepackage{microtype}      
\usepackage{lipsum}		
\usepackage{graphicx}
\usepackage{natbib}
\usepackage{doi}
\usepackage{amsmath} 
\usepackage{amssymb}
\usepackage{xcolor}

\title{Phase Stability and Superconductivity in Hydrogenated and Lithiated Janus GaXS$_2$ (X = Ga, In) Monolayers.}

\author{ \href{https://orcid.org/0009-0004-2196-8245}{\includegraphics[scale=0.06]{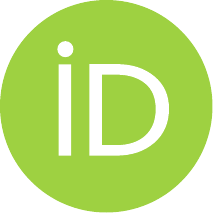}\hspace{1mm}Jakkapat Seeyangnok} \\
	Department of Physics\\
    Faculty of Science\\
	Chulalongkorn University\\
	Bangkok, Thailand \\
	\texttt{jakkapatjtp@gmail.com} \\
	\And
	\href{https://orcid.org/0000-0002-8450-7751}{\includegraphics[scale=0.06]{orcid.pdf}\hspace{1mm}Udomsilp Pinsook} \\
	Department of Physics\\
    Faculty of Science\\
	Chulalongkorn University\\
	Bangkok, Thailand \\
	\texttt{Udomsilp.P@Chula.ac.th} \\
}

\begin{document}
\maketitle

\begin{abstract}
Hydrogen and lithium functionalization of two-dimensional (2D) materials offers a promising route to enhance electronic properties and induce superconductivity. Here, we employ first-principles calculations to explore the phase stability and superconducting behavior of hydrogenated and lithiated Janus GaXS$_2$ (X = Ga, In) monolayers. Among Ga$_2$SH, Ga$_2$SLi, GaInSH, and GaInSLi, only the 2H-GaInSLi structure is dynamically, thermally, and mechanically stable, as confirmed by phonon dispersion, \textit{ab initio} molecular dynamics, and elastic constants. This monolayer adopts a hexagonal lattice, exhibits metallic behavior, and has a negative formation energy, suggesting experimental feasibility. Anisotropic Migdal–Eliashberg analysis reveals phonon-mediated superconductivity with a critical temperature ($T_c$) of 4.8 K. Notably, three distinct superconducting gaps emerge, linked to specific atomic orbitals and phonon modes. Electron doping of 0.2 $e$ per cell increases $T_c$ to nearly 6.2 K while maintaining the three-gap character. These results highlight the effectiveness of selective functionalization in engineering superconductivity and identify GaInSLi as a promising platform for next-generation multi-gap 2D superconducting devices.
\end{abstract}


	\section{Introduction}\label{sec1}
Janus transition-metal dichalcogenides (JTMDs) are a subclass of 2D materials characterized by out-of-plane asymmetry, arising from the presence of two different chalcogen elements on either side of a transition-metal layer. This broken mirror symmetry leads to tunable electronic, mechanical, and optical properties \cite{tang20222d,zhang2022janus,angeli2022twistronics,he2018two,yeh2020computational,yin2021recent}. Despite their absence in nature, 2D-JTMDs have been synthesized using advanced techniques such as selective epitaxy atomic replacement (SEAR), starting with Janus graphene in 2013 \cite{zhang2013janus} and expanding to MoSSe, WSSe, and PtSSe monolayers \cite{trivedi2020room,lu2017janus,sant2020synthesis}. 

Hydrogen-decorated 2D materials \cite{bekaert2019hydrogen,han2023high,liu2024three,seeyangnok2025high_npj2d,seeyangnok2025hydrogenation_nanoscale} and transition metal dichalcogenides (TMDs) have recently attracted significant attention due to their potential to exhibit superconductivity. Hydrogen substitution at chalcogen sites has led to the experimental realization of Janus MoSH monolayers \cite{lu2017janus}, with theoretical predictions of superconductivity up to $T_c = 27$~K in the 2H phase of MoSH reported in \cite{liu2022two}, followed by subsequent studies that further explored its electronic and superconducting properties \cite{ku2023ab,li2024machine}. A possible bilayer structure of MoSH was also investigated to assess the dimensional effects on its physical behavior \cite{pinsook2025superconductivity}. Beyond molybdenum-based compounds, similar studies have been conducted on other transition metals. For instance, tungsten-based TMD hydrides have been systematically analyzed, revealing superconducting behavior in certain configurations \cite{seeyangnok2024superconductivity,seeyangnok2024superconductivitywseh} along with the evidence of charge density wave (CDW) \cite{ku2023ab,qiao2024prediction,sui2025two}, while titanium-based systems have also been explored for their potential \cite{ul2024superconductivity}. However, hydrogen substitution does not universally induce superconductivity; in some materials, it leads to magnetic ordering as the ground state \cite{crsh_earth2025}. Moreover, in group IV transition metal systems, a delicate competition between magnetic and non-magnetic phases has been observed, with small energy differences suggesting possible coexistence or tunability under external perturbations \cite{seeyangnok2025competition}. Some of these systems have been reported to exhibit half-metallic ferromagnetism and magnetic ordering, making them promising candidates for spintronic applications.

Lithium decoration enhances the electronic properties of graphene and has been shown to induce superconductivity with a transition temperature around 8.1~K \cite{profeta2012phonon}, which was experimentally observed at 5.9~K using angle-resolved photoemission spectroscopy (ARPES) \cite{ludbrook2015evidence}. ARPES is a powerful technique to probe the quasiparticle dynamics driven by electron-phonon interactions. Since then, numerous lithium-decorated 2D materials have been explored, such as phosphorene \cite{haldar2017first}, SiB monolayers \cite{jiang2023lithium}, $\Psi$-graphene \cite{dewangan2023lithium}, lithium-decorated orthorhombic (o)-B$_2$X$_2$ monolayers \cite{benaddi2024lithium}, and Irida-graphene \cite{zhang2024li}. These findings establish lithium-functionalized 2D materials as promising candidates for next-generation electronics and superconducting technologies.

Inspired by these advancements, lithium substitution at the chalcogen site presents a novel direction to modify the surface chemistry and electronic properties of 2D-JTMDs \cite{singh2021review,chen2021interstitial,xie2024strong}. This approach introduces asymmetric charge distribution and potentially enhances electron-phonon coupling, offering new routes to high-performance 2D superconductors. These systems generally exhibit conventional BCS-type superconductivity, and in several cases, multigap behavior first recognized in MgB$_2$ \cite{souma2003origin} has been reported. In such materials, electron-phonon coupling from multiple electronic bands near the Fermi level can significantly enhance the superconducting transition temperature ($T_c$).

In this study, we propose and investigate a new Janus 2D material: GaInSLi, derived from GaInS$_2$~\cite{vu2021structural,betal2022strain,bikerouin2023solar} via lithium substitution on one chalcogen layer. This substitution results in a stable 2H hexagonal monolayer with broken mirror symmetry. We explore its electronic structure, vibrational properties, and superconducting behavior. The electron-phonon interaction is analyzed within the Migdal-Eliashberg theory \cite{bardeen1957microscopic,frohlich1950theory,migdal1958interaction,eliashberg1960interactions,nambu1960quasi,pinsook2024analytic}. This comprehensive investigation provides insights into the superconducting mechanism and multigap nature of GaInSLi, shedding light on its potential as a new 2D superconductor.

	\section{Methods}
The initial crystal structures were constructed using \textsc{VESTA}~\cite{momma2011vesta} within the trigonal space group \( P\overline{3}m1 \) (No.~156). Structural optimizations were carried out using the BFGS algorithm~\cite{BFGS,liu1989limited}, with full relaxation of atomic positions and lattice constants until the forces on each atom were below \(10^{-5}~\text{eV/\AA}\). Optimized norm-conserving Vanderbilt pseudopotentials~\cite{hamann2013optimized,schlipf2015optimization} were employed, with the generalized gradient approximation (GGA) in the Perdew-Burke-Ernzerhof (PBE) form~\cite{perdew1996generalized} used for the exchange-correlation functional. A plane-wave kinetic energy cutoff of 80~Ry and a charge density cutoff of 320~Ry were used throughout. The Brillouin zone was sampled with a Monkhorst-Pack $20 \times 20 \times 1$ k-point mesh~\cite{monkhorst1976special}, and Methfessel-Paxton smearing of 0.02~Ry was applied near the Fermi surface~\cite{methfessel1989high}. Fermi surface plots were visualized using \textsc{XCRYSDEN}~\cite{kokalj2003computer}. All first-principles calculations were performed using density functional theory (DFT) as implemented in the \textsc{Quantum ESPRESSO} (QE) package~\cite{giannozzi2009quantum,giannozzi2017advanced}. We investigated the possible magnetic configurations of GaInSLi, including ferromagnetic (FM) ordering, where the initial spin orientations of Ga and In atoms were aligned in the same direction, and antiferromagnetic (AFM) orderings. The AFM configurations considered were G-type (GAF), characterized by opposite spin orientations for neighboring atoms within the Ga plane and in the upper In plane, and A-type (AAF), where spins are aligned parallel within the Ga plane but antiparallel in the In plane.

Phonon dispersion relations and dynamical matrices were obtained via density functional perturbation theory (DFPT) using a $10 \times 10 \times 1$ q-point grid. The interaction between electrons and phonons leads to a finite lifetime for phonons, manifesting as a phonon linewidth given by:

\begin{equation}
\gamma_{\boldsymbol{q}\nu} = 2\pi\omega_{\boldsymbol{q}\nu}\sum_{nm}\sum_{\boldsymbol{k}} \left| g_{\boldsymbol{k}+\boldsymbol{q},\boldsymbol{k}}^{\boldsymbol{q}\nu,mn} \right|^2 \delta(\epsilon_{\boldsymbol{k}+\boldsymbol{q},m}-\epsilon_F) \delta(\epsilon_{\boldsymbol{k},n}-\epsilon_F),
\label{gammaphononlinewidths}
\end{equation}

where \( g_{\boldsymbol{k}+\boldsymbol{q},\boldsymbol{k}}^{\boldsymbol{q}\nu,mn} \) is the electron-phonon matrix element and \(\omega_{\boldsymbol{q}\nu}\) is the phonon frequency. The corresponding electron-phonon coupling constant for each mode is calculated via:

\begin{equation}\label{eqn:lambda_qv}
\lambda_{\boldsymbol{q}\nu} = \frac{\gamma_{\boldsymbol{q}\nu}}{\pi N(\epsilon_F) \omega_{\boldsymbol{q}\nu}^2},
\end{equation}

where \( N(\epsilon_F) \) is the electronic density of states at the Fermi level.

The Eliashberg spectral function, which encapsulates the energy-dependent electron-phonon interaction, is defined as:

\begin{equation}
\alpha^2 F(\omega) = \int \frac{d^3 q}{(2\pi)^3} \delta(\omega - \omega_{\boldsymbol{q}\nu}) \delta(\xi_{\boldsymbol{k}+\boldsymbol{q}}) \left| g_{\boldsymbol{k}+\boldsymbol{q},\boldsymbol{k}} \right|^2,
\label{eq:a2F}
\end{equation}

To accurately evaluate superconducting properties, including the total electron-phonon coupling constant \(\lambda\) and the Eliashberg spectral function \(\alpha^2 F(\omega)\), we utilized the Wannier-Fourier interpolation scheme implemented in the EPW code~\cite{noffsinger2010epw,ponce2016epw}, based on the formalism of Giustino \textit{et al.}~\cite{giustino2007electron,giustino2017electron}.

The anisotropic Migdal-Eliashberg equations were solved self-consistently to determine the superconducting gap function \(\Delta_{nk}(i\omega_j)\) and the renormalization factor \(Z_{nk}(i\omega_j)\) on the imaginary axis using Matsubara frequencies \(\omega_j = (2j+1)\pi T\):

\begin{equation}
Z_{nk}(i\omega_j) = 1 + \frac{\pi T}{N(\varepsilon_F)\omega_j} \sum_{mk'j'} \frac{\omega_{j'}}{\sqrt{\omega_{j'}^2 + \Delta_{mk'}^2(i\omega_{j'})}},
\label{eqn-ME1}
\end{equation}

\begin{eqnarray}
Z_{nk}(i\omega_j)\Delta_{nk}(i\omega_j) &= \frac{\pi T}{N(\varepsilon_F)} \sum_{mk'j'} \frac{\Delta_{mk'}(i\omega_{j'})}{\sqrt{\omega_{j'}^2 + \Delta_{mk'}^2(i\omega_{j'})}} \delta(\epsilon_{mk'} - \varepsilon_F) \nonumber \\
&\times \left[\lambda(nk, mk', \omega_j - \omega_{j'}) - \mu^*\right],
\label{eqn-ME2}
\end{eqnarray}

where \(\mu^*\) is the Morel-Anderson Coulomb pseudopotential, set to 0.1 in our calculations. The electron-phonon calculations were carried out using k- and q-point grids of \(240 \times 240 \times 1\) and \(120 \times 120 \times 1\), respectively. These dense grids ensured convergence of the \(\lambda\) and \(\alpha^2 F(\omega)\) values. The Fermi surface broadening was set to 0.180~eV, with a Matsubara frequency cutoff of 0.5~eV. Gaussian broadening was used to approximate the Dirac delta functions, with widths of 0.040~eV for electrons and 0.5~meV for phonons.

To assess the feasibility of synthesizing the material, we can investigate its energetic stability through both the cohesive ($E_{\text{coh}}$) and the formation energy ($E_{\text{form}}$) analyses using

\begin{equation}\label{eqn:cohesiveenergies}
    E = E_{\text{GaInSLi}} - E_{\text{Ga}} - E_{\text{In}} - E_{\text{S}} - E_{\text{Li}},
\end{equation}

where \(E_{\text{GaInSLi}}\) is the total energy of the monolayer. When calculating the cohesive energy ($E_{\text{coh}}$), the reference states of \(E_{\text{Ga}}, E_{\text{In}}, E_{\text{S}}, E_{\text{Li}}\) correspond to the energies of isolated Ga, In, S, and Li atoms, respectively. In contrast, when evaluating the formation energy ($E_{\text{form}}$), the reference states of \(E_{\text{Ga}}, E_{\text{In}}, E_{\text{S}}, E_{\text{Li}}\) are taken relative to their respective elemental bulk phases~\cite{zhang1991chemical,kirklin2015open}.

    To evaluate the mechanical robustness of Janus GaInSLi monolayer, we investigated its elastic behavior by computing the elastic constants appropriate for a 2D hexagonal lattice. The elastic constants \(C_{ij}\) are derived from the curvature of the total energy surface with respect to the applied strain, given by:
    
    \begin{equation}\label{eqn:cij}
    C_{ij} = \frac{1}{S_0} \frac{\partial^2 E}{\partial \epsilon_i \partial \epsilon_j}
    \end{equation}
    
    where \(S_0\) denotes the equilibrium area of the unit cell, and \(\epsilon_i\), \(\epsilon_j\) are the components of strain in the plane.
    
    In a 2D hexagonal structure, the independent in-plane elastic constants are \(C_{11}\) and \(C_{12}\), while the shear elastic constant is derived as \(C_{66} = \frac{C_{11} - C_{12}}{2}\). These constants govern the linear stress--strain relationship through the generalized Hooke’s law for two-dimensional systems:
    
    \begin{equation}\label{eqn:c11c22}
    \sigma
    =
    \begin{bmatrix}
    C_{11} & C_{12} & 0 \\
    C_{12} & C_{22} & 0 \\
    0 & 0 & C_{66}
    \end{bmatrix}
    \varepsilon.
    \end{equation}

	\section{Results and discussion}
	\subsection{Crystal structure}
	\begin{figure}[h!]
		\centering
		\includegraphics[width=14cm]{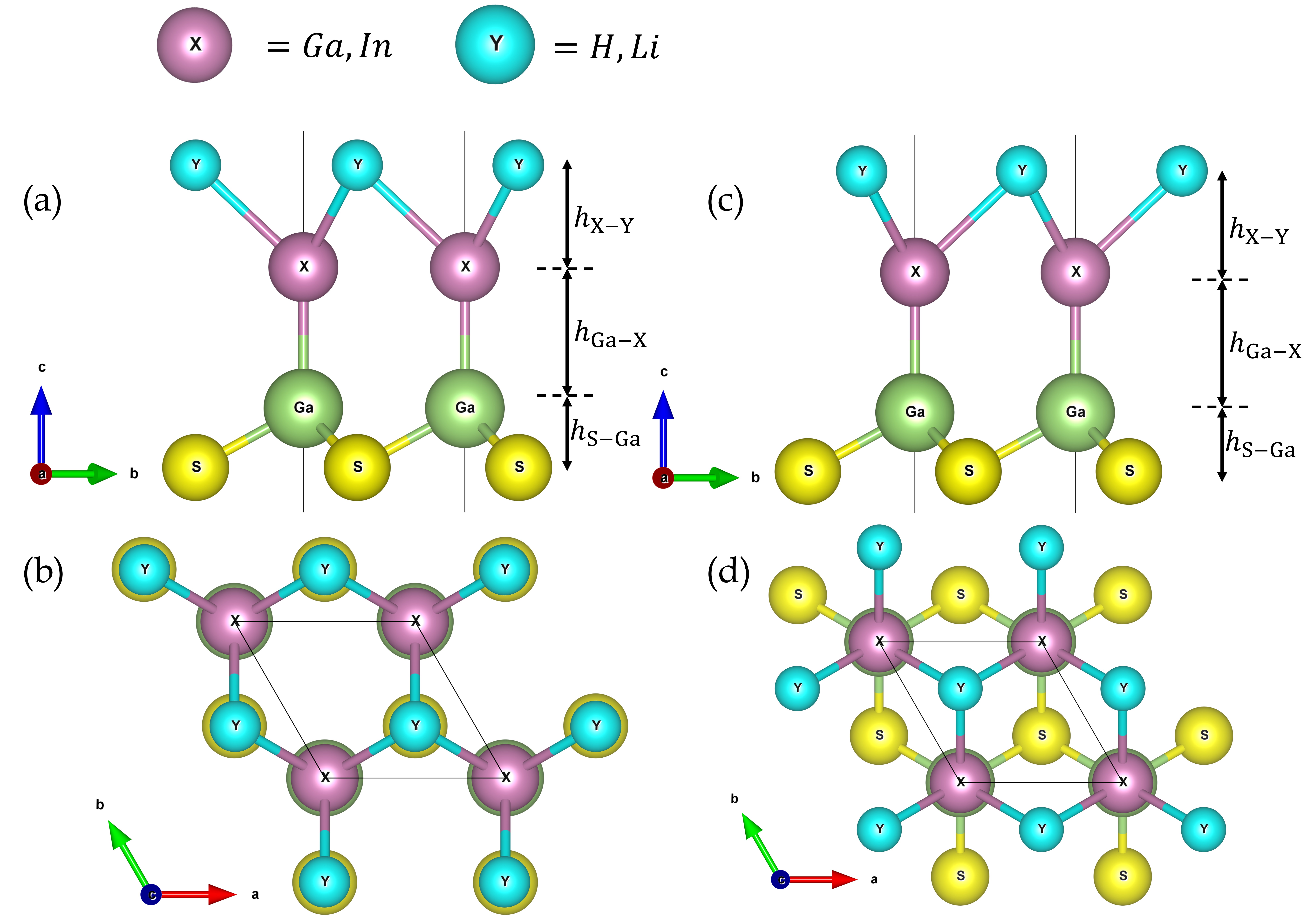}
		\caption{Top and side views of hydrogenated and lithiated Janus GaXS$_2$ (X = Ga, In; Y = H, Li) monolayers in the (a, b) 2H phase and (c, d) 1T phase. Panels (a) and (b) show the side and top views of the 2H configuration, while (c) and (d) present the corresponding views for the 1T configuration. The structural asymmetry arises from selective substitution of the top chalcogen layer by hydrogen or lithium atoms, forming a Janus-type structure. Interlayer vertical distances—$h_{\text{X--Y}}$, $h_{\text{Ga--X}}$, and $h_{\text{S--Ga}}$—are labeled for reference.}
		\label{fig:structure-GaXSY}
	\end{figure}

The periodically repeated hexagonal supercell of the Janus GaInSLi monolayer adopts the three-dimensional trigonal space group \( P\overline{3}m1 \) (No. 156), as illustrated in Figure~\ref{fig:structure-GaXSY}. In this structure, the transition metal atoms, Gallium (Ga) and Indium (In), occupy the Wyckoff position (0,0), while the chalcogen atom, sulfur (S), is located at the Wyckoff position \((1/3, 2/3)\) in the $xy$-plane with lattice constant $a$. The hydrogen (H) or lithium (Li) atom can, in principle, occupy either the \((1/3, 2/3)\) or \((2/3, 1/3)\) positions, corresponding to 2H and 1T stacking orders, respectively similar to the stacking polymorphs found in transition metal dichalcogenides (TMDs). We also investigated the possible magnetic configurations of GaInSLi, considering ferromagnetic (FM) and antiferromagnetic (AFM) orderings, including G-type (GAF) and A-type (AAF) antiferromagnetism. Among these, the non-magnetic (NM) metallic configuration is found to be the most stable. 

The energy computation of H- and Li-substituted Ga$_2$S$_2$ and GaInS$_2$ were investigated to find the most favorable configuration. For the H-substituted compounds, the 1T phase was found to be lower in energy than the 2H phase by 53.83 meV and 49.96 meV for Ga$_2$SH and GaInSH, respectively. In contrast, for the Li-substituted compounds, the 2H phase was lower in energy than the 1T phase by 15.63 meV and 13.04 meV for Ga$_2$SLi and GaInSLi, respectively. However, phonon calculations reveal that 1T-Ga$_2$SH and 1T-GaInSH are dynamically unstable due to imaginary phonon modes, as shown in Appendix~A, similar to the case of 2H-Ga$_2$SLi. Therefore, only the 2H-GaInSLi structure exhibits dynamical stability.

    \begin{table}[h!]
    \centering
    \caption{The table presents the lattice constant ($a$) and the interlayer distances: \( h_{\text{S--Ga}} \) (S to Ga layer), \( h_{\text{Ga--X}} \) (Ga to X layer), and \( h_{\text{X--Y}} \), all in units of~\AA, for the most energetically favorable configuration of each structure where X=Ga, In and Y=H, Li.}
    \begin{tabular}{|c|c|c|c|c|} 
     \hline
        2D materials & Lattice constant & \( h_{\text{S--Ga}} \) & \( h_{\text{Ga--X}} \) & \( h_{\text{X--Y}} \) \\ 
     \hline
     1T-Ga$_2$SH & 3.47 & 1.17 & 2.56 & 0.59 \\
     1T-GaInSH & 3.57 & 1.13 & 2.74 & 0.74 \\
     2H-Ga$_2$SLi & 3.51 & 1.15 & 2.50 & 1.73 \\
     2H-GaInSLi & 3.56 & 1.13 & 2.69 & 1.95 \\
  \hline
\end{tabular}
	\label{tab:lattice}
    \end{table}
    
\begin{figure}[h!]
		\centering
		\includegraphics[width=16cm]{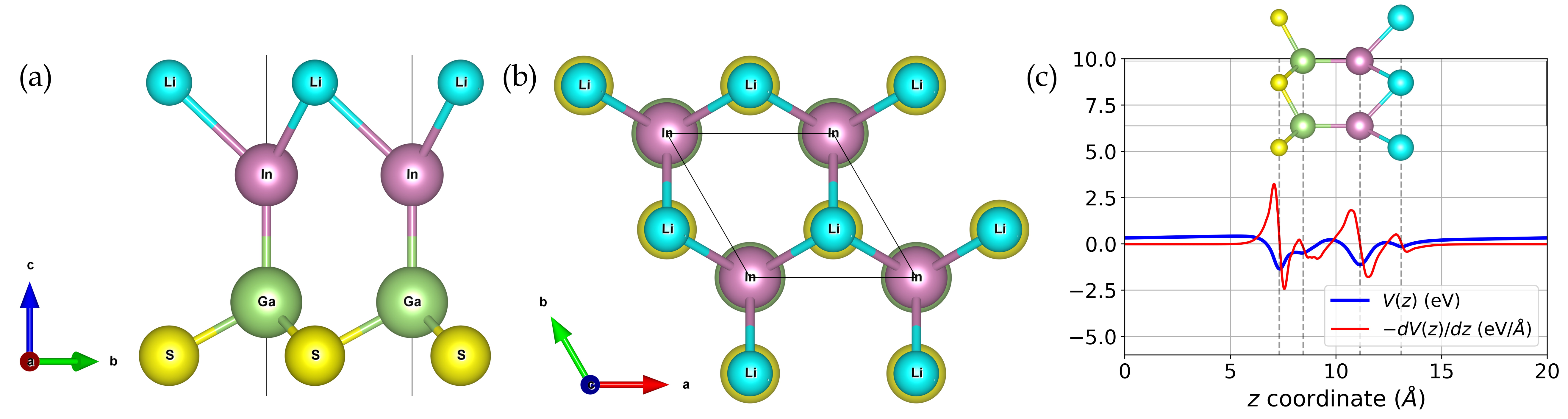}
		\caption{(a) Side view of the optimized crystal structure of the Janus GaInSLi monolayer, showing lithium atoms substituted on one side of the GaInS$_2$ host. The vertical distances between atomic layers—$h_\mathrm{Ga\text{-}S}$, $h_\mathrm{Ga\text{-}In}$, and $h_\mathrm{In\text{-}Li}$—are indicated. (b) Top view of the same structure highlighting the hexagonal arrangement of atoms and the coordination of lithium with indium atoms. The structure exhibits mirror symmetry breaking along the out-of-plane direction, characteristic of Janus monolayers. (c) illustrates the variation of the plane-averaged electrostatic potential \( V(z) \) (in eV) and the corresponding electrostatic field \( -dV(z)/dz \) (in eV/\AA) along the \( z \)-axis, where vertical dashed lines denote the positions of key atoms within the structure.
}
		\label{fig:structure}
	\end{figure}

The optimized structures corresponding to the most energetically favorable configurations are summarized in Table~\ref{tab:lattice}. For 2H-GaInSLi, the lattice constant is 3.56~\AA. Along the $z$-axis, the atomic configuration is characterized by the interplanar distances: \( h_{\text{In-Li}} = 1.95 \)~\AA{} (In to Li layer), \( h_{\text{Ga-In}} = 2.69 \)~\AA{} (Ga to In layer), and \( h_{\text{Ga-S}} = 1.13 \)~\AA (Ga to S layer) as shown in Figure~\ref{fig:structure} (a-b).

Finally, to investigate the presence of an intrinsic electric field arising from the asymmetry in GaInSLi, we analyzed the electrostatic potential profile. As shown in Figure~\ref{fig:structure}~(c), the asymmetric geometry results in an electrostatic potential difference between the top and bottom layers. This built-in potential difference, induced by the structural asymmetry, gives rise to an intrinsic electric field perpendicular to the layers. Although the asymmetric configuration introduces a electric potential across the vacuum region, the slope of the electrostatic potential confirms the absence of a significant net electric field across the monolayer.

    \subsection{Energetic stability}
To assess the feasibility of synthesizing the Janus GaInSLi monolayer, we investigated its energetic stability through both cohesive and formation energy analyses using Equation~\ref{eqn:cohesiveenergies}. The cohesive energy, calculated relative to the energies of isolated atoms, reflects the binding strength of atoms within the structure and is found to be \(-12.05~\text{eV}\) per formula unit, indicating strong thermodynamic stability of the monolayer. On the other hand, the formation energy (\(E_f\)) of the GaInSLi monolayer, calculated relative to its elemental bulk reference phases~\cite{zhang1991chemical,kirklin2015open}, is \(-1.15~\text{eV}\) per formula unit. Specifically, the energy of gallium was taken from its $\alpha$ phase, which crystallizes in the orthorhombic \textit{Cmce} space group~\cite{sondergaard2003electronic}; indium was taken in its body-centered tetragonal phase with \textit{I4/mmm} symmetry; sulfur was referenced from the orthorhombic $\alpha$-S\textsubscript{8} phase with \textit{Fddd} symmetry (space group No.~70)~\cite{fedyaeva2023stability}; and lithium was referenced from its body-centered cubic structure with \textit{Im}\={3}\textit{m} symmetry.

The significantly negative values of both the cohesive and formation energies suggest that the Janus GaInSLi monolayer is thermodynamically stable. These results imply that the synthesis of this structure may be feasible, potentially through conventional techniques such as chemical vapor deposition (CVD).

    \subsection{Thermal stability}  
    \begin{figure}[h!]
		\centering
		\includegraphics[width=16cm]{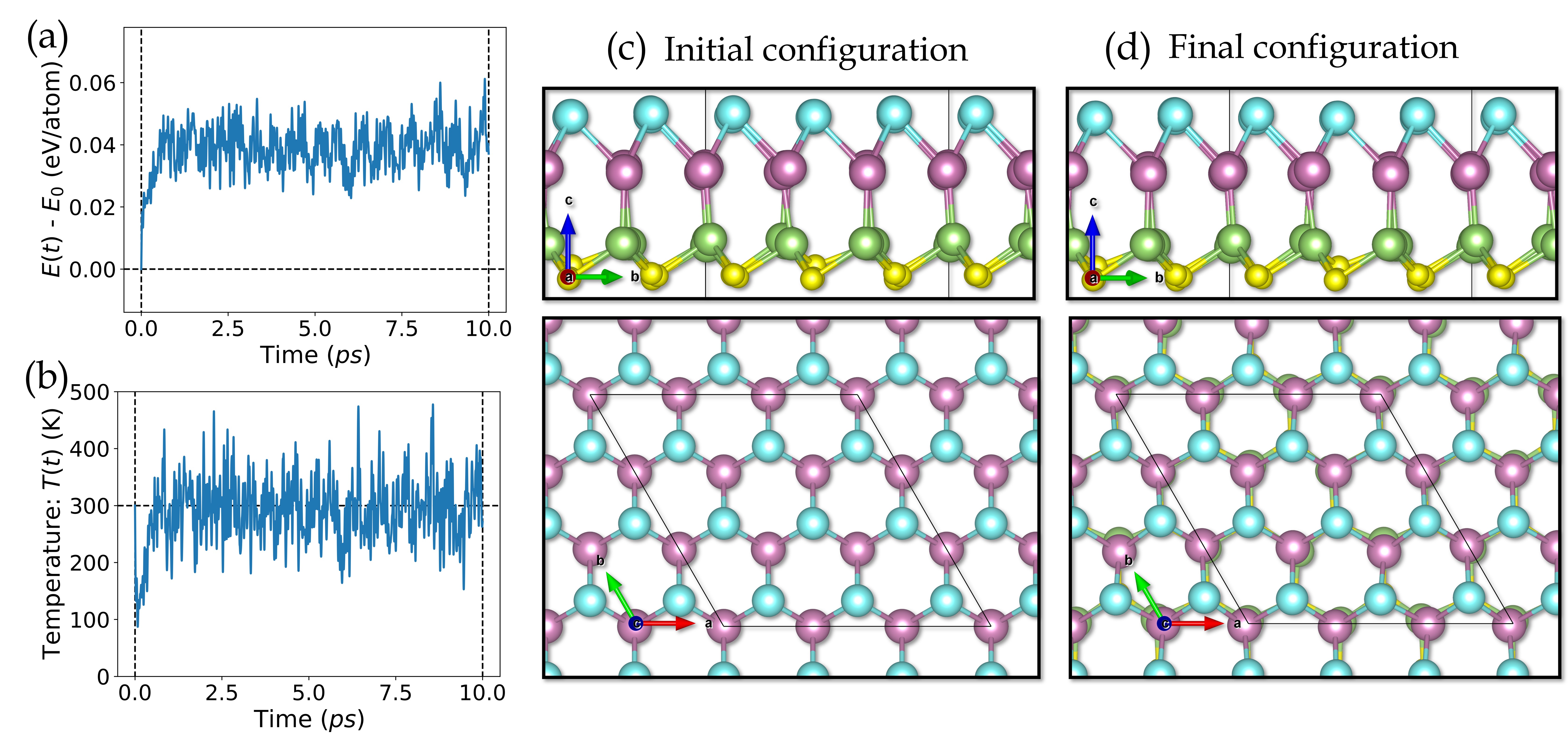}
		\caption{(a) Time evolution of the total energy fluctuation per atom, $E(t) - E_0$, during a 10 ps ab initio Born–Oppenheimer molecular dynamics (BOMD) simulation in the NVT ensemble at 300 K for a 3×3×1 supercell (36 atoms) of the Janus monolayer. (b) Corresponding temperature profile $T(t)$, showing stable thermal fluctuations around the target temperature. (c) Side and top views of the initial atomic configuration before simulation. (d) Side and top views of the final configuration after 10 ps, demonstrating the dynamic structural stability of the system.}
  	\label{fig:MD-plos}
	\end{figure}
   The thermal stability of the Janus GaInSLi monolayer was examined using \textit{ab initio} molecular dynamics (AIMD) simulations. Unlike phonon and elastic stability, which are derived from harmonic approximations, AIMD enables the assessment of finite-temperature behavior, such as structural integrity under thermal agitation or potential phase transitions. This method also helps identify possible surface reconstructions that may not be captured by zero-temperature calculations.

    A simulation were performed on a $3 \times 3 \times 1$ supercells (36 atoms) within the canonical (NVT) ensemble. The system was subjected to a finite-room temperature, and its evolution over time was monitored of 20,000 steps with time step ($dt$) of 0.5$fs$. The results show that the atomic arrangement remains essentially unchanged throughout the simulation, and the total energy fluctuates stably around the equilibrium value. These observations, illustrated in Figure~\ref{fig:MD-plos}, affirm that the GaInSLi monolayer maintains structural integrity at finite temperatures, complementing the phonon-based dynamical stability discussed later.

    \subsection{Mechanical stability} 
    We examined the mechanical stability of the Janus GaInSLi monolayer by computing the elastic constants characteristic of a two-dimensional hexagonal structure using Equations~\ref{eqn:cij} and \ref{eqn:c11c22}. Our calculations yield values of \(C_{11} = C_{22} = 3.36~\text{eV/\AA}\) and \(C_{12} = 1.41~\text{eV/\AA}\). These results satisfy the essential mechanical stability conditions for 2D systems: \(C_{11}C_{22} - C_{12}^2 > 0\) and \(C_{11}, C_{22}, C_{66} > 0\), as outlined in the work of Mouhat and Coudert~\cite{mouhat2014necessary}. This confirms that the GaInSLi monolayer is mechanically stable.
    
    In combination with phonon dispersion analysis indicating dynamical stability, \textit{ab initio} molecular dynamics revealing thermal robustness, and favorable energetic indicators such as a negative formation energy and significant cohesive energy, the GaInSLi monolayer emerges as a promising candidate for synthesis and practical application.

	
    	\subsection{Electronic properties}
   \begin{figure}[h!]
		\centering
		\includegraphics[width=13cm]{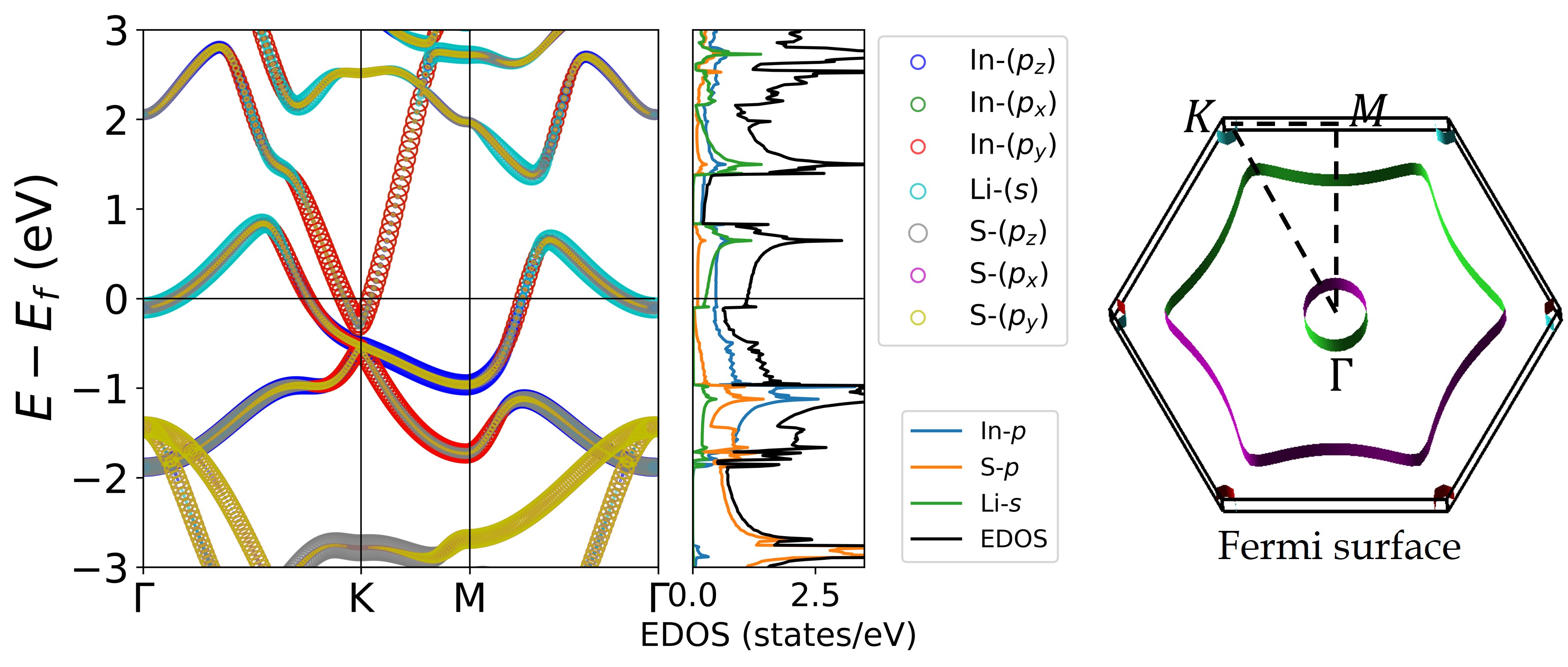}
		\caption{Figures show electronic structure and Fermi surface of the material. (Left) Orbital-resolved electronic band structure along the high-symmetry path $\Gamma$–K–M–$\Gamma$. The size and color of the markers represent the weight of atomic orbital contributions, including In-$p$, S-$p$, and Li-$s$ states. (Middle) Projected electronic density of states, showing contributions from In-$p$ (blue), S-$p$ (orange), and Li-$s$ (green) orbitals, along with the total electronic density of states (black). (Right) Fermi surface plotted in the two-dimensional Brillouin zone, exhibiting a central pocket around the $\Gamma$ point and an outer warped hexagonal contour, indicative of anisotropic multiband behavior.}
		\label{fig:electronics}
    \end{figure}    	
    The electronic structures are presented in Figure~\ref{fig:electronics}. These figures illustrate the characteristic behavior of electrons near the Fermi level, including the orbital-resolved electronic band structure, the total and orbital-projected electronic density of states, and the Fermi surface in the Brillouin zone. GaInSLi exhibits metallic behavior due to band crossings at the Fermi level, primarily originating from In-$p$, Li-$s$, and S-$p$ orbitals, in descending order of contribution. These band crossings occur near the $\Gamma$ and $K$ points, as well as along the $\Gamma$–$K$ and $\Gamma$–$M$ paths. As shown in Figure~\ref{fig:electronics}, these intersections give rise to three distinct Fermi surfaces. The orbital-resolved band structure reveals that the central Fermi surface around the $\Gamma$ point is dominated by the Li-$s$ orbital, while the Fermi surface around the $K$ point is mainly derived from the In-$p_y$ orbital. The remaining Fermi surface features, located between $\Gamma$–$K$ and $\Gamma$–$M$, arise from a mixture of orbitals, including contributions from both In-$p$ and S-$p$ states.

    The distinct orbital contributions to each Fermi surface suggest the possibility of different superconducting gap energies, potentially leading to orbital-selective pairing mechanisms for electrons located on different regions of the Fermi surface.
	\subsection{Phonons}
	\begin{figure}[h!]
		\centering
		\includegraphics[width=11cm]{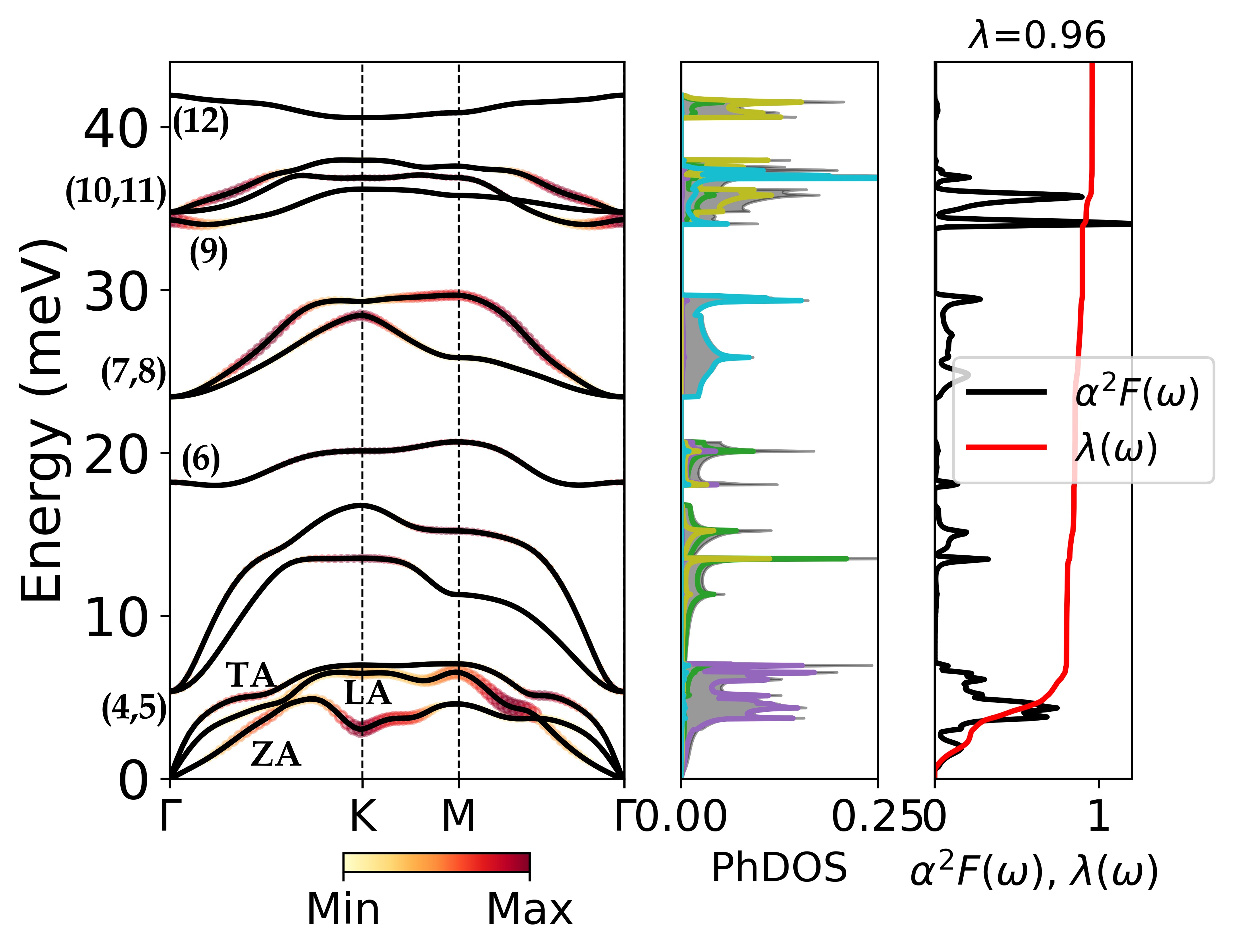}
		\caption{Figures show the phonon and electron-phonon coupling properties of GaInSLi. 
    Left panel: The electron-phonon coupling (EPC)-weighted phonon dispersion curves along high-symmetry directions in the Brillouin zone, with EPC strength visualized by a color gradient (from minimum to maximum). 
    Middle panel: Phonon density of states (PhDOS) with atom-projected contributions, showing the vibrational modes of different elements: purple (In), green (Ga), yellow (S), and cyan (Li). 
    Right panel: Eliashberg spectral function \( \alpha^2F(\omega) \) (black curve) and the integrated electron-phonon coupling constant \( \lambda(\omega) \) (red curve). The total EPC constant is \( \lambda = 0.96 \), indicating moderately strong electron-phonon interactions relevant for conventional superconductivity.
    }
		\label{fig:phonons}
	\end{figure}
    The dynamical stability of GaInSLi is confirmed by its positive-definite phonon spectrum, as shown in Figure~\ref{fig:phonons}. At the $\Gamma$ point, the phonon modes correspond to the $C_{3v}$ ($3m$) point group symmetry, consisting of three acoustic and six optical branches. The acoustic modes include the in-plane longitudinal acoustic (LA), in-plane transverse acoustic (TA), and out-of-plane flexural (ZA) vibrations. Notably, the ZA mode exhibits phonon softening near the $K$ point, manifesting as a dip in the dispersion. This softening enhances the electron-phonon coupling at the corresponding wavevector $\boldsymbol{q}$, as evidenced by the electron-phonon coupling-weighted phonon dispersion also shown in Figure~\ref{fig:phonons}.
    
\begin{figure}[h!]
		\centering
		\includegraphics[width=15cm]{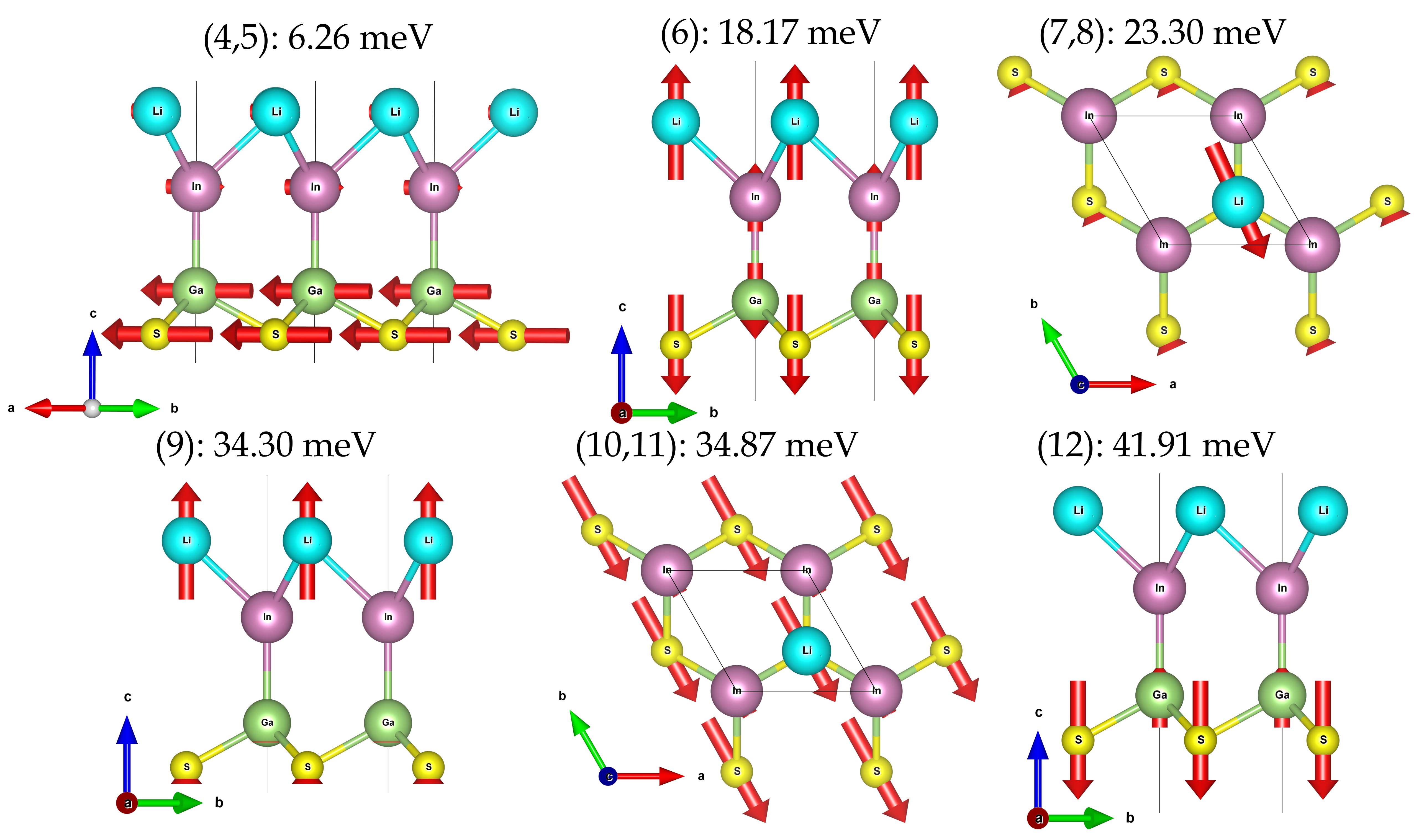}
		\caption{Figures show the visualization of the vibrational modes at the $\Gamma$ point corresponding to the optical phonon modes listed in Table~\ref{tab:phonon-eigenvalues}. Each mode illustrates the relative atomic displacements within the unit cell, providing insight into the character and symmetry of the phonon modes.}
		\label{fig:eigenvectors}
	\end{figure}

    The optical phonon branches can be classified into two symmetry subgroups: $E$ and $A_1$, as detailed in Table~\ref{tab:phonon-eigenvalues}. Bands (4,5) at 6.26~meV arise from in-plane vibrations of Ga and S atoms. Band (6) at 18.17~meV is dominated by out-of-plane vibrations involving Ga, S, In, and Li atoms. Bands (7,8) at 23.30~meV are primarily due to in-plane vibrations of Li, while band (9) at 34.30~meV corresponds to its out-of-plane motion. Bands (10,11) at 34.87~meV originate from in-plane vibrations of S, and the highest mode, band (12) at 41.91~meV, is attributed to out-of-plane vibrations of S. These vibrational patterns are further illustrated in Figure~\ref{fig:eigenvectors}. Due to the lack of inversion symmetry in the crystal structure, all vibrational modes are both Raman and infrared (IR) active. A comprehensive summary of the phonon eigenmodes is provided in Table~\ref{tab:phonon-eigenvalues}.
    
    \begin{table}[ht]
    \centering
	\caption{\label{tab:phonon-eigenvalues}
summarizes all the important information of the phonon modes, such as band numbers, subgroups, eigenvectors, infrared (I) or Raman (R) active, the frequencies (meV) at the $\Gamma$ point of GaInSLi monolayer.}
	   \begin{tabular}{lclclclclcl}
        \hline 
    Bands $\nu$ & Subgroup & Eigenvector & Active & energy (meV) \\
	   \hline 
		4,5 & $E~L_3$ & \text{In-plane Ga, S} & I+R & 6.26 \\
		6 & $A_1~L_1$ & \text{Out-plane Ga, S, In, Li} & I+R & 18.17 \\
		7,8 & $E~L_3$ & \text{In-plane Li} & I+R & 23.30 \\
		9 & $A_1~L_1$ & \text{Out-plane Li}  & I+R & 34.30 \\
        10,11 & $E~L_3$ & \text{In-plane S}  & I+R & 34.87 \\
        12 & $A_1~L_1$ & \text{Out-plane S}  & I+R & 41.91 \\
            \hline 
	   \end{tabular}
	\end{table}
\subsection{Superconductivity}
    
    \begin{figure}[h!]
		\centering
		\includegraphics[width=16cm]{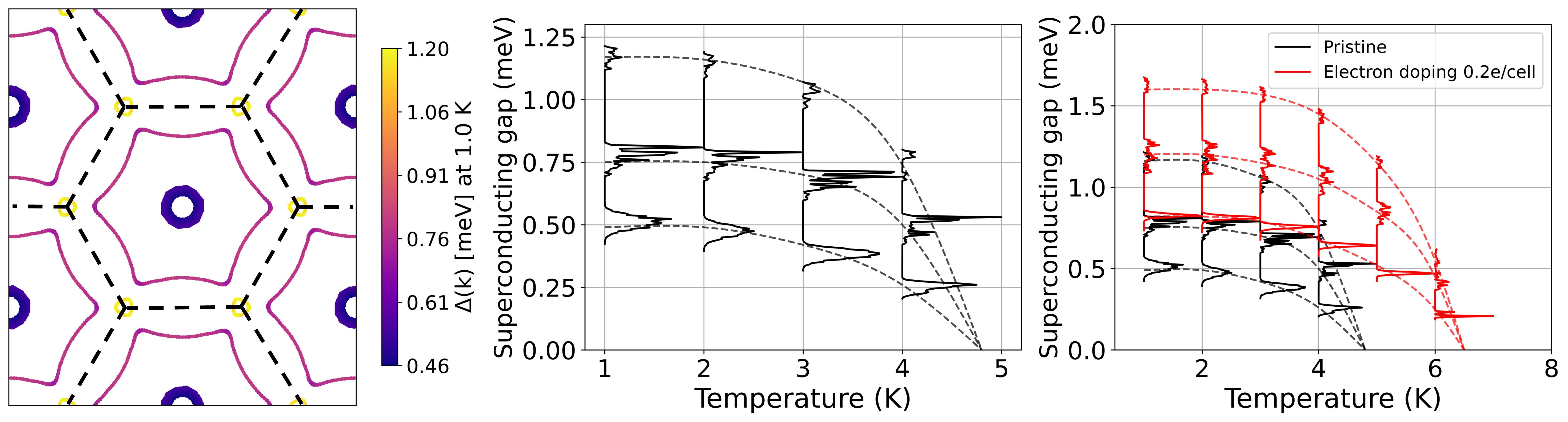}
		\caption{Left: Fermi surface of the material with a colormap indicating the momentum-resolved superconducting gap $\Delta(\mathbf{k})$ at 1.0 K. The dashed black lines denote the Brillouin zone boundaries. The superconducting gap exhibits significant anisotropy ranging from 0.46 to 1.20 meV. Middle and Right: Temperature dependence of the superconducting gap for two different cases: pristine (black) and electron-doped with $0.2e/\text{cell}$ (red). Solid black and red lines represent the calculated superconducting gaps, while dashed lines show interpolations of each gap. Electron doping enhances both the superconducting gap and critical temperature from 4.8 K to 6.2 K where there is no consistent numerical solution to the Migdal–Eliashberg equations. The presence of multiple gap branches indicates multiband superconductivity.}
		\label{fig:epw-gap}
	\end{figure}

    Superconductivity driven by electron-phonon coupling can be characterized by the wavevector-resolved electron-phonon coupling strength $\lambda_{n\boldsymbol{q}}$, as shown in the EPC-weighted phonon dispersion in Figure~\ref{fig:phonons}. A significant contribution to the electron-phonon coupling ($\lambda$) arises from the softening of the out-of-plane flexural mode (ZA) at the $K$ point and the in-plane longitudinal mode (LA) near the $M$ point. Together, the three acoustic modes account for approximately 81\% of the average electron-phonon coupling, $\lambda = 0.96$, which is obtained by integrating the Eliashberg spectral function $\alpha^2 F(\omega)$, as shown in Figure~\ref{fig:phonons}.
 
    The superconducting transition temperature is determined by the closing of superconducting gaps, calculated using the two nonlinear coupled anisotropic Migdal-Eliashberg equations (Equations~\ref{eqn-ME1} and~\ref{eqn-ME2}). As shown in Figure~\ref{fig:epw-gap}, GaInSLi exhibits three-gap superconductivity. The superconducting gap contour at 1.0~K reveals three distinct regions, each represented by different colors, indicating the presence of three superconducting gaps, as shown in Figure~\ref{fig:epw-gap} (left). The superconducting gaps are divided into three regions with values of approximately 1.15~meV, 0.75~meV, and 0.50~meV, providing clear evidence of three-gap superconductivity. This observation is further confirmed by the temperature dependence of the superconducting gaps, as shown in Figure~\ref{fig:epw-gap} (right). Other 2D materials exhibiting three superconducting gaps have also been reported, such as InB$_2$ and InB$_4$~\cite{wang2021three}, LiBCH~\cite{liu2024three}, and NaB$_2$C$_2$~\cite{meng2024unraveling}.
    
    In relation to the electronic structure, the largest superconducting gap appears near the $K$ point and is primarily associated with the In-$p_y$ orbital. This large gap is attributed to strong electron-phonon coupling (EPC) near the $K$ point, particularly involving the flexural (ZA) phonon mode of In, as illustrated in Figure~\ref{fig:phonons}. An intermediate superconducting gap emerges between the $\Gamma$ point and the zone boundaries, resulting from the presence of a Fermi surface in this region that originates from a mixture of multiple orbitals. The smallest superconducting gap is observed around the $\Gamma$ point and is mainly contributed by the Li-$s$ orbital, driven by EPC near $\Gamma$ associated with mode 9, corresponding to the out-of-plane vibration of Li atoms, as shown in Figure~\ref{fig:phonons}. These three distinct orbital contributions across different regions of the Fermi surface give rise to the three-gap superconducting behavior observed in GaInSLi.
    
    With increasing temperature, the superconducting gaps $\Delta(K)$ progressively diminish and fully close at 4.8~K, marking the superconducting critical temperature. This behavior confirms the intrinsic three-gap superconducting nature of GaInSLi. We also investigated the modulation of superconducting gaps in 2H-GaInSLi through doping, as reflected by the increased electronic density of states. Our results show that electron doping of 0.2 $e$ per cell raises $T_c$ to nearly 6.2 K while preserving the three-gap character with $\lambda = 1.55$. These findings demonstrate that the orbital-selective multi-gap behavior of GaInSLi enables tunable superconductivity under doping.

    \subsection{General discussion}

     Two-dimensional (2D) lithium-decorated materials have attracted significant interest since lithium-decorated graphene (LiC$_6$)~\cite{profeta2012phonon,ludbrook2015evidence} is a well-established benchmark, exhibiting superconducting gap anisotropy~\cite{zheng2016first} and a transition around 6~K. In contrast, 2H-GaInSLi has a Janus monolayer structure with top-bottom asymmetry, resulting in three-gap superconductivity rather than single-gap anisotropy. Although its superconducting critical temperature ($\sim$4.8~K) is slightly lower than that of LiC$_6$, its orbital-selective multigap behavior allows tunability under strain, doping, or external fields, which is not accessible in LiC$_6$. For example, the electron doping of $0.2e/\text{cell}$ increases $T_c$ to approximately 6.2~K.

    Although our calculations utilize the conventional adiabatic Migdal–Eliashberg framework (\(\omega_{D}/E_{F} \ll 1\)), we note that 2H–GaInSLi may exhibit non-adiabatic behavior, similar to other two-dimensional hexagonal superconductors where relatively narrow conduction bands play a crucial role. Prior studies on lithium-decorated graphene have revealed signatures of non-adiabatic superconductivity~\cite{szczȩsniak2019signatures,hu2022nonadiabatic}. The degree of non-adiabaticity can be qualitatively estimated by comparing the characteristic energy scales of electrons and phonons through the ratio \(\omega_{D}/E_{F}\). For LiC\(_6\)~\cite{zheng2016first}, this ratio is \(\omega_{D}/E_{F} \approx 0.15\) corresponding to ~\cite{szczȩsniak2019signatures}, based on a C-\(p\)-derived electron pocket with \(E_{F}=1.2\)~eV and the highest C vibrational mode at 180~meV. In the case of 2H–GaInSLi, we obtain \(\omega_{D}/E_{F} \approx 0.21{-}0.32\), derived from a Li-\(s\)-dominated electron pocket with \(E_{F}=0.11\)~eV and the corresponding Li vibrational modes: (7,8) at 23.30~meV and (9) at 34.87~meV. The similarity of these energy scales highlights the breakdown of the strict validity of the Migdal approximation. Taken together, these features suggest that non-adiabatic effects may contribute to enhancing or modifying the superconducting state in 2H–GaInSLi, thereby motivating future studies employing non-adiabatic formalisms.
    
    Following these pioneering works, recent studies have extended the exploration of lithium-decorated systems to Janus transition-metal chalcogenides (TMCs). For instance, the hexagonal Janus MoSLi and MoSeLi monolayers have been predicted to exhibit metallic behavior and phonon-mediated superconductivity with critical temperatures of about 14.8~K~\cite{xie2024strong} and 4.5~K~\cite{moseliseeyangnok}, respectively. Anisotropic Migdal--Eliashberg calculations further revealed that both MoSLi and MoSeLi host two-gap superconductivity, highlighting the potential of Li-decorated Janus structures as promising multigap superconductors. In MoSLi, strong electron--phonon coupling driven by Mo-$d_{z^2}$ orbitals and low-frequency phonon modes is responsible for its relatively high $T_c$. Within this context, our study on GaInSLi introduces an even more intriguing scenario: the presence of three distinct superconducting gaps, making it an exceptional candidate among Li-decorated Janus systems. Such multi-gap characteristics are also observed in a variety of other 2D superconductors, including $n$-doped graphene~\cite{margine2014two}, AlB$_2$-derived thin films~\cite{zhao2019two}, trilayer LiB$_2$C$_2$~\cite{gao2020strong}, monolayer LiBC~\cite{modak2021prediction}, NaB$_2$C$_2$ and Na$_2$B$_3$C$_3$~\cite{meng2024unraveling}, LiBCH~\cite{liu2024three}, and various transition-metal-based systems such as MoSH~\cite{liu2022two}, and InB$_2$ and InB$_4$~\cite{wang2021three}.
  
	\section{Conclusion}
    In summary, we have systematically investigated the structural, electronic, dynamical, and superconducting properties of the Janus GaInSLi monolayer. First-principles calculations reveal that GaInSLi adopts a stable trigonal structure with space group \( P\overline{3}m1 \), and exhibits a negative formation energy of \(-1.15~\text{eV}\) per formula unit. \textit{Ab initio} molecular dynamics simulations confirm the structural integrity of GaInSLi at finite temperatures, indicating thermodynamic stability. Additionally, the mechanical stability is verified by satisfying the elastic stability criteria, while the absence of imaginary phonon modes confirms its dynamical stability.

    Electronic structure analysis reveals metallic behavior, with bands crossing the Fermi level predominantly contributed by In-$p$, Li-$s$, and S-$p$ orbitals. Superconductivity in GaInSLi is driven by strong electron-phonon interactions, as characterized by the wavevector-resolved electron-phonon coupling strength $\lambda_{n\boldsymbol{q}}$. Notably, significant contributions arise from the softening of the acoustic phonon branches—specifically, the out-of-plane flexural (ZA) mode at the $K$ point and the in-plane longitudinal (LA) mode near the $M$ point. These three acoustic modes together contribute approximately 81\% to the total electron-phonon coupling strength, yielding an average coupling constant of $\lambda = 0.96$, as obtained by integrating the Eliashberg spectral function $\alpha^2F(\omega)$.
    
    Solving the anisotropic Migdal-Eliashberg equations reveals the emergence of three distinct superconducting gaps. These gaps are associated with different regions of the Fermi surface: the largest near the $K$ point (In-$p_y$ orbital), a moderate one between $\Gamma$, $K$, and $M$ points (mixed orbitals), and the smallest near the $\Gamma$ point (Li-$s$ orbital). As temperature increases, the superconducting gaps $\Delta(k)$ continuously diminish, vanishing at the critical temperature of 4.8~K. Electron doping of 0.2 $e$ per cell increases $T_c$ to nearly 6.2 K while maintaining the three-gap character.
    
    This intrinsic three-gap superconductivity places GaInSLi among a unique class of multigap superconductors. Our findings suggest GaInSLi as a promising platform for exploring orbital-selective superconductivity and tunable multigap phenomena in two-dimensional materials, potentially guiding future experimental realizations.

    \section*{Data Availability}
    The data that support the findings of this study are available from the corresponding
    authors upon reasonable request.
    
    \section*{Code Availability}
    The first-principles DFT calculations were performed using the open-source Quantum ESPRESSO package, available at \url{https://www.quantum-espresso.org}, along with pseudopotentials from the Quantum ESPRESSO pseudopotential library at \url{https://pseudopotentials.quantum-espresso.org/}. Electron-phonon coupling and related properties were computed using the EPW code, available at \url{https://epw-code.org/}.

    \section*{Acknowledgements}
	This research project is supported by the Second Century Fund (C2F), Chulalongkorn University. We acknowledge the supporting computing infrastructure provided by NSTDA, CU, CUAASC, NSRF via PMUB [B05F650021, B37G660013] (Thailand). URL:www.e-science.in.th.

    \section*{Author Contributions}
    Jakkapat Seeyangnok performed all of the calculations, analysed the results, wrote the first draft manuscript, and coordinated the project. Udomsilp Pinsook analysed the results and wrote the manuscript.

    \section*{Conflict  of Interests}
    The authors declare no competing financial or non-financial interests.

    \section*{Appendix A: Unstable Janus GaInSH, Ga$_2$SH, and Ga$_2$SLi monoalyer}\label{appendixa}
        \begin{figure}[ht]
		\centering  
    	\includegraphics[width=15cm]{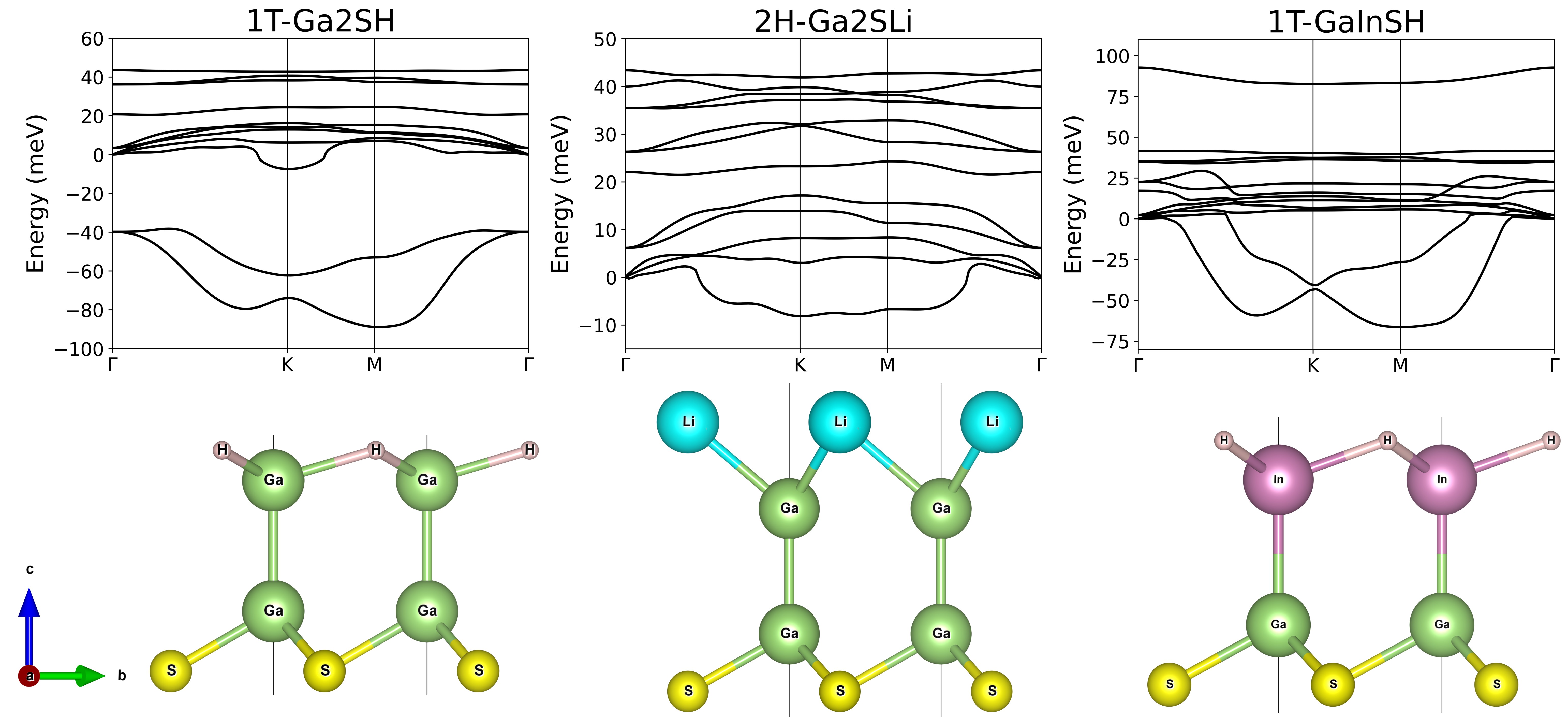}
		\caption{The figures present a comparison between electronic bands computed using standard DFT calculations (black) and those obtained via maximally-localized Wannier functions (dashed red). Additionally, the figures illustrate the Fermi surface around the Fermi level, which is essential for computing the electron-phonon interaction when solving the anisotropic Migdal-Eliashberg equations.}
		\label{fig:unstable-phonons}
    \end{figure}  
    
    We investigated various forms of H- and Li-substituted Ga$_2$S$_2$ and GaInS$_2$. For the H-substituted compounds, the 1T phase was energetically more favorable than the 2H phase by 53.84 meV and 49.94 meV for Ga$_2$S$_2$H and GaInS$_2$H, respectively. In contrast, for the Li-substituted counterparts, the 2H phase was more stable, with energy differences of 15.62 meV and 9.57 meV for Ga$_2$S$_2$Li and GaInS$_2$Li, respectively. To guide future studies and avoid unnecessary computational efforts, we highlight that phonon calculations reveal dynamical instabilities—indicated by imaginary phonon modes—in 1T-Ga$_2$SH, 1T-GaInSH, and 2H-Ga$_2$SLi, as shown in Figure~\ref{fig:unstable-phonons}. Among all configurations, only 2H-GaInSLi is found to be dynamically stable.

    \section*{Appendix B: Wannier interpolation}
    \begin{figure}[ht]
		\centering  
    	\includegraphics[width=7cm]{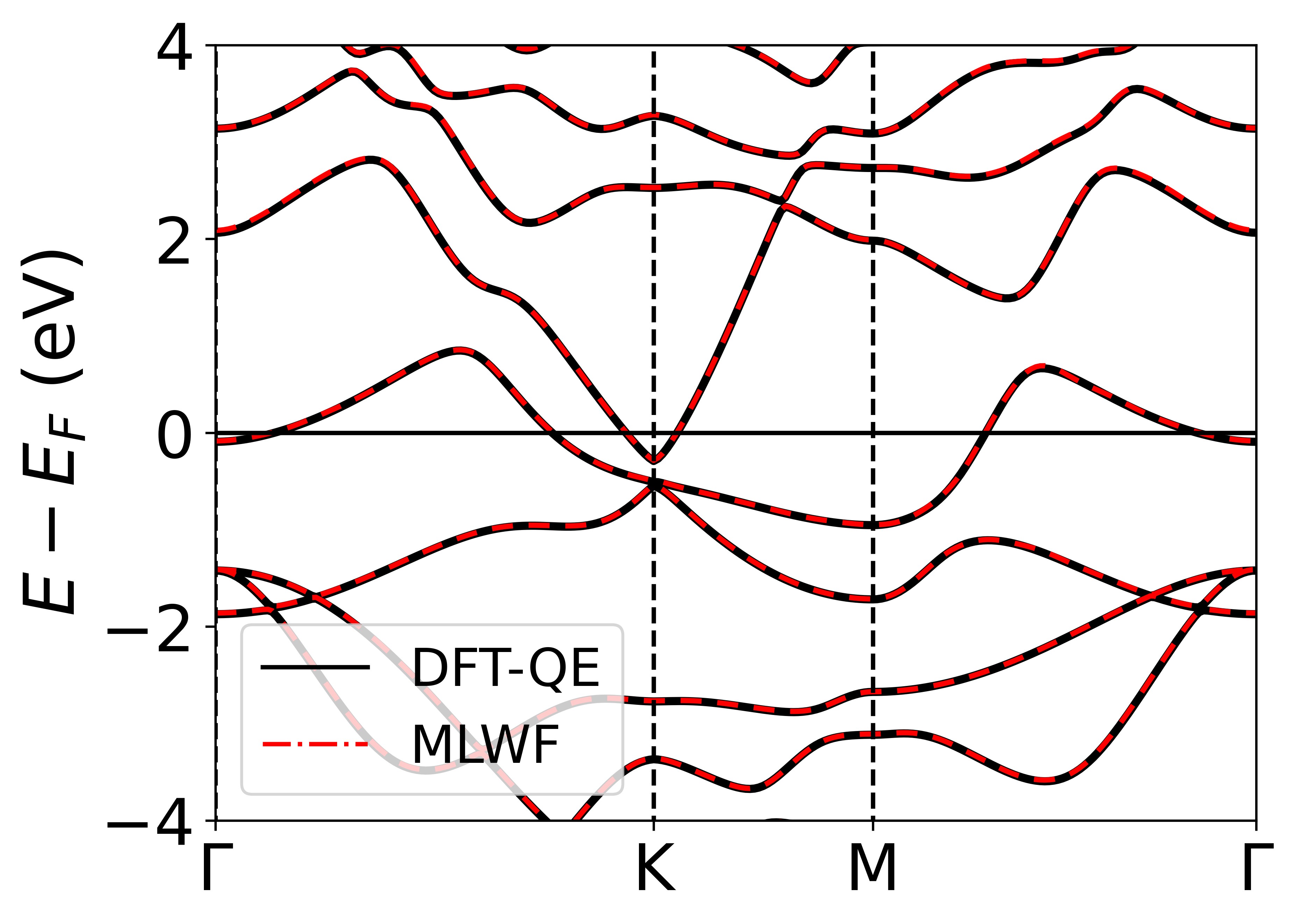}
		\caption{The figures present a comparison between electronic bands computed using standard DFT calculations (black) and those obtained via maximally-localized Wannier functions (dashed red). Additionally, the figures illustrate the Fermi surface around the Fermi level, which is essential for computing the electron-phonon interaction when solving the anisotropic Migdal-Eliashberg equations.}
		\label{fig:epw}
    \end{figure}  
    The maximally-localized Wannier functions (MLWFs), computed using the EPW package, are presented in Figure~\ref{fig:epw}. These MLWFs serve as an essential basis for accurately interpolating the electronic structure and electron-phonon matrix elements across the Brillouin zone. Their localization ensures efficient and reliable evaluation of the electron-phonon interaction, which is crucial for subsequent calculations of the Eliashberg spectral function and superconducting properties.
 
	\section*{References}
    \bibliographystyle{unsrt} 
	\bibliography{references}

\end{document}